\documentclass[twocolumn,aps,showpacs]{revtex4}

\usepackage{graphicx}
\usepackage{epsfig}
\usepackage{dcolumn}
\usepackage{amsfonts}
\usepackage{amsmath}
\usepackage{amssymb}
\usepackage{bm}

\begin{document}

\title{Transport on Directed Percolation Clusters}

\author{Hans-Karl Janssen}
\author{Olaf Stenull}
\affiliation{
Institut f\"{u}r Theoretische Physik 
III\\Heinrich-Heine-Universit\"{a}t\\Universit\"{a}tsstra{\ss}e 1\\
40225 D\"{u}sseldorf\\
Germany
}

\date{\today} 

\begin{abstract}
We study random lattice networks consisting of resistor like and diode like bonds. For investigating the transport properties of these 
random resistor diode networks we introduce a field theoretic Hamiltonian amenable to renormalization group analysis. We focus on 
the average two-port resistance at the transition from the nonpercolating to the directed percolating phase and calculate the 
corresponding resistance exponent $\phi$ to two-loop order. Moreover, we determine the backbone dimension $D_B$ of directed 
percolation clusters to two-loop order. We obtain a scaling relation for $D_B$ that is in agreement with well known scaling arguments.
\end{abstract}

\maketitle

\noindent
Percolation\cite{bunde_havlin_91_etc} is a leading paradigm for disorder. It provides an intuitively appealing and transparent model 
of the irregular geometry which occurs in disordered systems. Moreover, it is a prototype of a phase transition. Though the usual 
isotropic percolation (IP) has attracted most attention, directed percolation (DP)\cite{hinrichsen_2000} is a sexy model for study as 
well. DP shows many qualitatively new features not appearing in IP. DP is perhaps the simplest model resulting in branching 
self-affine objects. It has many potential applications, including fluid flow through porous media under gravity, hopping conductivity 
in a strong electric field\cite{vanLien_shklovskii_81}, crack propagation\cite{kertez_vicsek_80}, and the propagation of surfaces at 
depinning transitions\cite{depinning}. Furthermore, it is related to epidemics with a bias\cite{grassberger_85} and self-organized 
critical models\cite{soc}. While the transport properties of IP have been studied 
extensively\cite{sammlung,stenull_janssen_oerding_99,janssen_stenull_oerding_99,janssen_stenull_99,stenull_2000,stenull_janssen_2000a}, 
relatively little is known about transport in DP. The transport properties of DP have not been addressed hitherto by using sophisticated 
analytic methods like renormalized field theory.

A model which captures both, IP and DP, is the random resistor diode network (RDN) introduced by 
Redner\cite{red_81&82a,red_83,perc}. A RDN consists of a $d$-dimensional hypercubic lattice in which nearest-neighbor sites are 
connected by a resistor, a positive diode (conducting only in a distinguished direction), a negative diode (conducting only opposite 
to the distinguished direction), or an insulator with respective probabilities $p$, $p_{+}$, $p_{-}$, and $q=1-p-p_{+}-p_{-}$. In the 
three dimensional phase diagram (pictured as a tetrahedron spanned by the four probabilities) one finds a nonpercolating and three 
percolating phases. The percolating phases are isotropic, positively directed, or negatively directed. Between the phases there are 
surfaces of continuous transitions. All four phases meet along a multicritical line, where $0\leq r:=p_{+}=p_{-}\leq 1/2$ and 
$p=p_{c}(r)$. On the entire multicritical line, i.e., independently of $r$, one finds the scaling properties of usual isotropic percolation 
($r=0$). For the crossover from IP to DP see, e.g., Ref.\cite{janssen_stenull_2000}.

In this letter we focus on the vicinity of the critical surface separating the nonpercolating and the positively directed phase. Here, 
typical clusters are anisotropic and they are characterized by two different correlation lengths: $\xi_{\parallel}$ (parallel to the 
distinguished direction) and $\xi_\perp$ (perpendicular to it). As one approaches the critical surface, the two correlation lengths 
diverge with the exponents $\nu_\parallel$ and $\nu_\perp$ of the DP universality class.

In the first part of this letter we study the average resistance between two connected sites $x$ and $x^\prime$ when an external 
current $I$ is injected at $x$ and withdrawn at $x^\prime$. We choose ${\rm{\bf n}} = 1/\sqrt{d} \left( 1, \dots , 1 \right)$ for the 
distinguished direction. We assume that the bonds $\underline{b}_{\langle i,j \rangle}$ between two nearest neighboring sites $i$ and 
$j$ are directed so that $\underline{b}_{\langle i,j \rangle} \cdot {\rm{\bf n}} > 0$. The directed bonds obey the non-linear Ohm's law
\begin{eqnarray}
\sigma_{i,j} \left( V_j - V_i \right) \left[ V_j - V_i \right] = I_{i,j}\ ,
\end{eqnarray}
where $V_i$ is the potential at site $i$ and $I_{i,j}$ denotes the current flowing from $j$ to $i$. The bond conductances $\sigma_{i,j}$ 
are random variables taking on the values $\sigma$, $\sigma \theta \left( V \right)$, $\sigma \theta \left( -V \right)$, and $0$ with 
respective probabilities $p$, $p_+$, $p_-$, and $q$. $\sigma$ is a positive constant and $\theta$ denotes the Heaviside function. 
Note that the diodes are idealized: under forward-bias voltage they behave as ``ohmic'' resistors whereas they are insulating under 
backward-bias voltage.

A central role in our theory is played by the power
\begin{eqnarray}
\label{power}
P \left( \left\{ V \right\} \right) = \sum_{< i, j>} \sigma_{i,j} \left( V_j -V_i \right) 
\left[ V_j -V_i \right]^2
\end{eqnarray} 
dissipated on the network. The sum in Eq.~(\ref{power}) is taken over all bonds of the lattice. Following an idea by 
Stephen\cite{stephen_78} and its generalization to networks of nonlinear resistors by Harris\cite{harris_87} we exploit correlation 
functions of $\psi_\lambda \left( x \right) = \exp \left( i \lambda V_x \right)$ as generating functions of the resistance $R \left( x, 
x^\prime \right)$ between $x$ and $x^\prime$. Note that $\lambda = i I$ is an imaginary current. With help of the saddle point method 
(the integration is not Gaussian due to the $\theta$ functions) we find
\begin{eqnarray}
\label{generator}
&&\left\langle \psi_\lambda \left( x \right) \psi_{-\lambda} \left( x^\prime \right) \right\rangle 
\nonumber \\
&&  = \, 
\frac{1}{Z} \int \prod_i dV_i \, \exp \left[ - \frac{1}{2} P \left( \left\{ V \right\} \right) + i \lambda \left( V_x -V_{x^\prime} \right) \right]
 \nonumber \\
&& \propto \, \exp \left[ - \frac{\lambda^2}{2} R \left( x, x^\prime \right) \right] \ ,
\end{eqnarray}
provided that the condition $I^2 \gg \sigma$ holds. $Z$ in Eq.~(\ref{generator}) stands for the usual normalization. 

We are interested in the average $\langle \ldots \rangle_C$ of $R$ over all diluted lattice configurations which we will denote by 
$M_R$. Hence we switch to $D$-fold replicated voltages $V_i \to \vec{V_i} = \left( V_i^{(1)}, \cdots , V_i^{(D)} \right)$ and imaginary 
currents $\lambda_i \to \vec{\lambda_i} = \left( \lambda_i^{(1)}, \cdots , \lambda_i^{(D)} \right)$. The replication procedure induces the 
effective Hamiltonian
\begin{eqnarray}
H_{\mbox{\scriptsize{rep}}} &=&  - \ln \left\langle  \exp \left[ - \frac{1}{2} P \left( \left\{ \vec{V} \right\} \right) \right] \right\rangle_C \ .
\end{eqnarray}
For technical reasons\cite{harris_lubensky_87} we switch to discretized voltages $\vec{\theta}$ and currents $\vec{\lambda}$ taking 
values on a discrete $D$-dimensional torus. For the saddle point method to be reliable we work near the limit when all the 
components of $\vec{\lambda}$ are equal and continue to large imaginary values. Accordingly we set\cite{harris_87} 
$\lambda^{(\alpha )} = i \lambda_0 + \xi^{(\alpha )}$ with real positive $\lambda_0$ and $\xi^{(\alpha )}$, $\sum_{\alpha =1}^D 
\xi^{(\alpha )} = 0$, and impose the conditions $\lambda_0^{2} \ll D^{-1}$ and $\vec{\xi}^2 \ll 1$.

To refine $H_{\mbox{\scriptsize{rep}}}$ towards a field theoretic Hamiltonian, we expand $H_{\mbox{\scriptsize{rep}}}$ in terms of 
$\psi_{\vec{\lambda}} \left( x \right)$. The steps are analogous to those in Ref.\cite{harris_87} and are skipped here for briefness. The 
so obtained expression is converted into a Landau-Ginzburg-Wilson-type functional
\begin{eqnarray}
\label{hamiltonian}
{\mathcal{H}} &=& \int d^dx \bigg\{ \frac{1}{2} \sum_{\vec{\lambda} \neq \vec{0}} \psi_{-\vec{\lambda}} \left( {\rm{\bf x}} \right) \Big[  
\tau - \nabla^2 + w \vec{\lambda}^2 
\nonumber \\
&&+ \, \left( \theta \left( \lambda_0 \right) - \theta \left( -\lambda_0 \right) \right) {\rm{\bf v}} \cdot 
\nabla \big] \psi_{\vec{\lambda}} \left( {\rm{\bf x}} \right)
\nonumber \\
& & + \, \frac{g}{6} \sum_{\vec{\lambda}, \vec{\lambda}^\prime  , \vec{\lambda} + \vec{\lambda}^\prime \neq \vec{0}} 
\psi_{-\vec{\lambda}} \left( {\rm{\bf x}} \right) \psi_{-\vec{\lambda}^\prime} \left( {\rm{\bf x}} \right) \psi_{\vec{\lambda} + 
\vec{\lambda}^\prime} \left( {\rm{\bf x}} \right) \bigg\}
\end{eqnarray}
by applying the usual coarse graining procedure. The parameter $\tau$ specifies the ``distance'' from the critical surface under 
consideration. The vector ${\rm{\bf v}}$ lies in the distinguished direction, ${\rm{\bf v}} = v {\rm{\bf n}}$. $\tau$ and $v$ depend on 
the three probabilities $p$, $p_+$, and $p_-$. $w$ is the coarse grained analog of $\sigma^{-1}$. In the limit $w\to 0$ our 
Hamiltonian ${\mathcal{H}}$ describes the usual purely geometric DP. Indeed ${\mathcal{H}}$ leads for $w\to 0$ to exactly the same 
perturbation series as obtained in\cite{cardy_sugar_80,janssen_81,janssen_2000}. 

We proceed with standard methods of field theory\cite{amit_zinn-justin} and perform a diagrammatic perturbation calculation up to 
two-loop order. As in our previous work on transport in 
IP\cite{stenull_janssen_oerding_99,janssen_stenull_oerding_99,janssen_stenull_99,stenull_2000,stenull_janssen_2000a}, the 
principle propagator consists of an conducting and an insulating part. Hence, the principle Feynman diagrams can be decomposed 
into conducting diagrams consisting of conducting and insulating propagators. These conducting diagrams can be interpreted as 
being directed networks themselves. This real-world interpretation leads to a substantial simplification of the actual calculation. 
Instead of carrying out tedious summations over loop currents, we just have to determine the total resistance of the conducting 
diagrams. The remaining steps in calculating the diagrams are well known from the field theory of 
DP\cite{cardy_sugar_80,janssen_81,janssen_2000}.

Renormalization group analysis provides us with the scaling behavior of the correlation function
\begin{eqnarray}
\label{corrfkt}
&&\left\langle \psi_{\vec{\lambda}} \left( {\rm{\bf 0}} \right) \psi_{-\vec{\lambda}} \left( {\rm{\bf x}} \right) \right\rangle_{\mathcal{H}}
\nonumber \\ 
&& = \, 
\left| {\rm{\bf x}}_\perp \right|^{1-d-\eta} \, f_1 \left(  \frac{x_\parallel}{\left| {\rm{\bf x}}_\perp \right|^z} \right) 
\nonumber \\ 
&& \times \,
\bigg\{ 1 + w 
\vec{\lambda}^2 \left| {\rm{\bf x}}_\perp \right|^{\phi /\nu_\perp} f_{w,1} \left(  \frac{x_\parallel}{\left| {\rm{\bf x}}_\perp \right|^z} \right) + 
\cdots \bigg\} 
\nonumber \\
&& = \, 
x_\parallel^{(1-d-\eta )/z} \, f_2 \left(  \frac{\left| {\rm{\bf x}}_\perp \right|^z}{x_\parallel} \right) 
\nonumber \\ 
&& \times \,
\bigg\{ 1 + w \vec{\lambda}^2 
x_\parallel^{\phi /\nu_\parallel} f_{w,2} \left(  \frac{\left| {\rm{\bf x}}_\perp \right|^z}{x_\parallel} \right) + \cdots \bigg\} \ ,
\end{eqnarray}
where $\eta$, $\nu_\perp$, and $z= \nu_\parallel /\nu_\perp$ are the critical exponents for DP known to second order in 
$\epsilon$-expansion\cite{janssen_81,janssen_2000}. $f_1$, $f_2$, $f_{w,1}$, and $f_{w,2}$ are scaling functions. In 
Eq.~(\ref{corrfkt}) we introduced the resistance exponent $\phi$. Exploiting the fact that the correlation function in Eq.~(\ref{corrfkt}) is 
a generating function for $M_R$ (cf.\ Eq.~(\ref{generator})) we deduce that $M_R \sim \left| x_\parallel \right|^{\phi /\nu_\parallel}$ if 
measured parallel to the distinguished direction. For measurements in other directions it is appropriate to choose a length scale $L$ 
and to express the longitudinal and the transverse coordinates in terms of $L$: $\left| {\rm{\bf x}}_\perp \right| \sim L$ and $x_\parallel 
\sim L^z$. With this choice the scaling function $f_{w,1}$ reduces to a constant and we obtain $M_R \sim L^{\phi /\nu_\perp}$. For 
the resistance exponent we find in $\epsilon$-expansion
\begin{eqnarray}
\label{resPhi}
\phi = 1 + \frac{\epsilon}{24}  + \frac{151 - 314 \ln \left( 4/3 \right)}{6912} \, \epsilon^2 + {\sl O} \left( \epsilon^3 \right) \ ,
\end{eqnarray}
where $\epsilon =5-d$. Note that $\phi$ is larger than the corresponding resistance exponent for the random resistor network 
(RRN)\cite{lubensky_wang_85,stenull_janssen_oerding_99}. This is intuitively plausible since long tortuous paths that contribute to 
the macroscopic conductance in RRN are suppressed in DP.

Now we compare Eq.~(\ref{resPhi}) to the few numerical results available in the literature. We are not aware of any numerical results 
for $\phi$ itself. However, Redner and Mueller\cite{redner_mueller_82} determined the conductivity exponent $t = \phi + 
(d-1)\nu_\perp + \nu_\parallel$ in two dimensions by Monte Carlo simulations: $t (d=2) = 0.6 \pm 0.10$. Arora {\em et 
al}.\cite{arora&co_83} did analogue and numerical simulations leading to $t = 0.73 \pm 0.10$. Another value for comparison is $t 
\approx 0.7$ obtained by Redner\cite{red_83} from a real space renormalization group calculation. Crudely evaluating the 
$\epsilon$-expansion of $t$ for small spatial dimensions leads inevitably to poor quantitative predictions. Therefore it is appropriate to 
improve the expansion by incorporating rigorously known features. By carrying out a rational approximation which takes into account 
that $t (d=1) =0$ we obtain the interpolation formula
\begin{eqnarray}
\label{tAppr}
t \approx  \left( 1 - \frac{\epsilon}{4} \right) \left( 2 + 0.2083 \, \epsilon + 0.0604 \, \epsilon^2 \right) \ ,
\end{eqnarray}
which leads to $t (d=2)\approx 0.8$.

The second part of this letter is devoted to the backbone dimension in DP. The backbone between two sites $x$ and $x^\prime$ is 
defined, apart from unimportant Wheatstone-bridge-type configurations, as the union of all bonds carrying current when $I$ is 
inserted at $x$ and withdrawn at $x^\prime$. The average number of bonds belonging to the backbone is referred to as its mass 
$M_B$. For the purpose of calculating the backbone dimension $D_B$ we assume that the bonds obey a generalized Ohm's law
\begin{eqnarray}
\sigma_{i,j} \left( V_j - V_i \right) \left[ V_j - V_i \right] \left[ V_j - V_i \right]^{1/r-1} = I_{i,j}\ .
\end{eqnarray}
The parameter $r$ measures the nonlinearity of the bond. $\sigma_{i,j} \left( V_j - V_i \right)$ takes on the same values as in the first 
part. The field theoretic Hamiltonian for the generalized RDN is given by Eq.~(\ref{hamiltonian}) with $w \vec{\lambda}^2$ replaced by 
$- w_r \sum_{\alpha =a}^D \left( -i \lambda^{(\alpha )} \right)^{r+1}$.

In the following we are going to use that the average resistance of the generalized RDN, $M_{R_r}$, and the backbone mass are 
related via $M_B \sim \lim_{r \to -1^+}M_{R_r}$ (see, e.g., Ref.~\cite{janssen_stenull_99}). A two-loop calculation analogous to that 
in part one reveals that the coupling proportional to $w_r$ does not require an individual renormalization in the limit ${r \to -1^+}$. As 
a consequence we obtain
\begin{eqnarray}
\label{SkalenRel}
\lim_{r \to -1^+} \phi_r /\nu_\perp = z - \eta \ ,
\end{eqnarray}
at least to second order in $\epsilon$. From the generalized version of Eq.~(\ref{corrfkt}) we deduce that the backbone mass scales 
as
\begin{eqnarray}
\label{massScaling}
M_B = \left| {\rm{\bf x}}_\perp \right|^{\phi_{-1} /\nu_\perp} f_{w,1} \left(  \frac{x_\parallel}{\left| {\rm{\bf x}}_\perp \right|^z} \right) \ .
\end{eqnarray}

For self-affine objects the notion of fractal dimension is less straightforward than for self-similar objects. To determine the fractal 
dimension of the DP backbone one considers a $d-1$-dimensional hyper-plane with an orientation perpendicular to $x_\parallel$. 
The cut is a self similar object with the fractal dimension
\begin{eqnarray}
\label{cutDim}
d_{\mbox{\scriptsize{cut}}} = D_B - 1\ ,
\end{eqnarray}
where $D_B$ is the local fractal dimension of the backbone\cite{kertez_vicsek_94}. According to Eq.~(\ref{massScaling}) the mass of 
the cut scales like
\begin{eqnarray}
\label{cutScaling}
M_{\mbox{\scriptsize{cut}}} = \left| {\rm{\bf x}}_\perp \right|^{\phi_{-1} /\nu_\perp} x_\parallel^{-1} f_{w,1} \left(  \frac{x_\parallel}{\left| 
{\rm{\bf x}}_\perp \right|^z} \right) \ .
\end{eqnarray}
By choosing once more $\left| {\rm{\bf x}}_\perp \right| \sim L$ and $x_\parallel \sim L^z$ we find that $M_{\mbox{\scriptsize{cut}}} \sim 
L^{\phi_{-1} /\nu_\perp - z}$. This leads via Eqs.~(\ref{cutDim}) and (\ref{SkalenRel}) to
\begin{eqnarray}
\label{DBfoermelchen}
D_B = 1 + \phi_{-1} /\nu_\perp - z = 1 - \eta = d - 2\beta / \nu_\perp \ ,
\end{eqnarray}
where $\beta = \nu_\perp (d-1+\eta )/2$ is the DP order parameter exponent known to second order in 
$\epsilon$\cite{janssen_81,janssen_2000}.

Equation~(\ref{DBfoermelchen}) is in agreement with scaling arguments\cite{hede&co_91} yielding that the fractal dimension of the 
transverse cut through a DP cluster with local dimension $d_f$ is $d_f -1 = d-1 -\beta / \nu_\perp$. The analogous cut through the 
backbone can be viewed as the intersection of the cut through the cluster and the clusters backward oriented 
pendant\cite{red_83,arora&co_83}. Hence, the codimension of the backbone cut is twice the codimension $\beta / \nu_\perp$ of the 
cluster cut, which leads again to Eq.~(\ref{DBfoermelchen}).

We conclude with a few remarks. Our approach gives Eq.~(\ref{DBfoermelchen}) perturbatively to second order in $\epsilon$, while 
the scaling arguments leading to Eq.~(\ref{DBfoermelchen}) are exact. Hence, Eq.~(\ref{DBfoermelchen}) has a manifestation in the 
renormalization group framework in form of some Ward identity. The fact that $w_{-1}$ renormalizes trivially to two-loop order is 
reminiscent of this Ward identity. It is an interesting issue for future work to identify the Ward identity and its underlying symmetry. Our 
result for the resistance exponent $\phi$ is for dimensions close to five the most accurate analytic estimates that we know of. In two 
dimensions our results show reasonable agreement with the known numerical results. It is certainly desirable to have more and firmer 
numerical data for comparison with our analytic results, in particular in three dimensions. We hope that this letter triggers further 
simulations of transport in DP.  

We acknowledge support by the Sonderforschungsbereich 237 ``Unordnung und gro{\ss}e Fluktuationen'' of the Deutsche 
Forschungsgemeinschaft. We thank S. Redner for bringing our attention to Ref.~\cite{arora&co_83} and for making us aware of an 
error in our initial calculation. Finally, we thank an anonymous referee for pointing out subtleties in the definition of $D_B$.


\end{document}